\documentclass[preprint]{ptephy_v1}

\preprintnumber{XXXX-XXXX} 

\usepackage{graphicx}  
\usepackage{dcolumn}  
\usepackage{bm}        
\usepackage{amssymb}
\usepackage{amsmath,mathtools,cases, empheq} 
\usepackage{epstopdf} 
\usepackage{amsfonts,float}
\usepackage{fancyhdr}
\usepackage{pifont}

\usepackage[hidelinks]{hyperref} 
\usepackage{bbm}
\usepackage{tensor}
\usepackage{mathrsfs}
\usepackage{xcolor}
\usepackage{physics}

		\newcommand{\der}[3]{\frac{d^#3 #1}{d#2^#3}}
		\newcommand{\deruno}[2]{\frac{d#1}{d#2}}

\begin{document}

\title{Casimir energy through transfer operators for sine-Gordon backgrounds}

\author{Luc{\'i}a Santamar{\'i}a-Sanz}
\affil{Departamento de F{\'i}sica Te{\'o}rica, At{\'o}mica y {\'O}ptica, Universidad de Valladolid, 47011 Valladolid, Spain.\\Departamento de F{\'i}sica, Universidad de Burgos, 09001 Burgos, Spain \email{lssanz@ubu.es}}

\begin{abstract}
The quantum vacuum interaction energy  between a pair of semitransparent two-dimensional plates represented by Dirac delta potentials and its first derivative,  embedded in the topological background of a sine-Gordon kink,  is studied through an extension of the \textit{TGTG}-formula (developped by O. Kenneth and I. Klich in the scattering approach).  Quantum vacuum oscillations around the sine-Gordon kink solutions are interpreted as a quantum scalar field theory in the spacetime of a domain wall.   Moreover, the relation between the phase shift and the density of states (the well-known Dashen-Hasslacher-Neveu formula) is also exploited to characterize the quantum vacuum energy.
\end{abstract}

\subjectindex{xxxx, xxx}

\maketitle

\section{Introduction}
\label{sec:intro}
The vacuum state of an arbitrary Quantum Field Theory (QFT) is the state of minimum energy of the Hamiltonian. This state is not empty but contains electromagnetic waves and  infinite pairs of virtual short-lived particles and antiparticles \cite{Colemanbook}. These pairs annihilate each other very quickly in accordance with the Heisenberg energy-time uncertainty principle. Nevertheless, when some objects are introduced into the space, sometimes nearby particles collide with them and get reflected in such a way that they do not combine again with their antipaticles. Quantum forces may thus appear in the system as a result of the presence of these frontiers. The physical properties of the vacuum state and the vacuum energy show a strong dependence on the type of boundaries. One of the most important boundary phenomenon is the Casimir effect \cite{Casimir1948}, in which the presence of two parallel, uncharged and conducting plates restrics the modes of the fluctuations of the electromagnetic field between them. Vacuum polarization occurs and a finite force between both plates appears. This effect, predicted in 1948 by H.B.G Casimir, was experimentally measured for the first time in 1958 by M.J. Sparnaay \cite{Sparnaay1957}. Since then, many studies have been performed for bodies with different geometries and materials \cite{Bordagbook, Emig2017review}. This effect has numerous applications in nanoelectronic devices \cite{Fosco2009, Allen2005}, in absorption phenomena in carbon nanotubes \cite{Klimchitskaya2008}, and in inflation process \cite{Mamaevbook},  among many others.

A crucial point of the two-plate theory is that the vacuum energy is ultraviolet divergent.  It needs to be regularized and renormalized in order to obtain the finite contribution related to the interaction betwen objects. There are several approaches to achieve this goal. For instance, it is worth mentioning:
\begin{itemize}
\item computing the 00-component of the energy-momentum tensor in terms of Green's function and the scattering data and then subtracting the first terms in its Born expansion \cite{Graham2002},
\item calculating the transfer operator to apply the \textit{TGTG}-formula \cite{KK2008},
\item or using zeta functions, complex integrals and heat traces \cite{Kirstenbook, Vassilevich2003, Elizaldebook1}.
\end{itemize} The $TGTG$-formula allows one to compute the quantum vacuum interaction energy between two disjoint objects represented by a smooth classical background in a flat spacetime.  It is a formalism based on transition operators (more precisely the so called Lippmann-Schwinger \textit{T}-operator) and Green's functions.  It has been used for instance to compute the vacuum interaction energy between two sine-Gordon kinks \cite{Bordag2012},  and between two plates mimicked by $\delta \delta^\prime$ potentials \cite{Guilarte2015}, like the ones that will be discussed in the following sections.  The major advantage of using the \textit{TGTG}-formalism instead of the other ones aforementioned is that only the scattering problem for the background potential together with one single object is necessary to compute the vacuum interaction energy between the pair of bodies. This could be a crucial factor whenever the scattering problem for the complete potential with two objects in a classical background is hard to be solved. Furthermore, using the \textit{TGTG}-formula significantly reduces  the complexity of the analytical and numerical computation.  For both reasons, it is the method to be used in the present work.

The main objective here will be finding a generalization of the \textit{TGTG}-formula to study the quantum vacuum interaction energy between a pair of two-dimensional homogeneous plates located\footnote{The position four-vector will be expressed from now on as $x^\mu=(t,\vec{x}_\parallel, z) \in \mathbb{R}^{1,3}$. Notice that $\vec{x}_\parallel \in \mathbb{R}^2$. Likewise, the four-momentum will be $K^\mu=(E, \vec{k}_\parallel, k)$. Here, $z$ and $k$ are the position  and the momentum coordinates in the direction orthogonal to the surfaces of the plates. }  at $z=a,b$, and immersed in a weak curved background potential. This background is constituted by a sine-Gordon kink centred at the origin of the direction orthogonal to the plates.  Thus,  the background potential will be the P\"oschl-Teller (PT) one:
\begin{eqnarray}
V_{PT}(z)=-2\sech^2 z = -\frac{2}{\cosh^2 z}=  -\frac{8}{\left(e^z+e^{-z}\right)^2}.
\end{eqnarray} 
It models the propagation of mesons moving in a sine-Gordon kink background \cite{Vachaspatibook,  Mantonbook,  Rajaramanbook, Guilarte2011}.  Notice that kinks in $1+1$ dimensions can be embedded into a $3+1$ dimensional theory as solutions which are independent of all but one spatial direction. These types of solutions with finite energy per unit area are known as domain walls \cite{Mantonbook}.  They were produced in the early stages of the Universe by the spontaneous breaking of discrete symmetries \cite{Press1989}. The propagation of mesons around this topological defect will be the focus in this article.  

Sine-Gordon kinks have attracted much attention up to date.  Previous works concerning the $00$-component of the energy momentum tensor $T_{\mu\nu}$ in a system with a kink in the real line, and the scattering problem of two delta potentials simetrically placed around a kink can be found in \cite{Cavero2010, Guilarte2011}.  Moreover,  the one-soliton sine-Gordon solution can be used to construct (locally) a metric describing a curved spacetime of a black hole \cite{GegenbergPLB1997},  to generate 4D Einstein gravity solutions \cite{VacaruJHEP2001},  and to provide braneworld scenarios \cite{BazeiaEPJC2017}.  In this last case,  the scalar field acts as a source of gravity around the brane.  So far, much is made of quantum forces in different curved backgrounds and in gravitational contexts by Saharian \cite{Saharian2004},  Bordag \cite{Munoz-Castaneda:2014lia},  Milton and Fulling \cite{Milton_2008},  Elizalde and Odintsov \cite{PhysRevD.67.063515},  among others.  There are relevant differences between studying QFT for flat metrics and doing it for those that characterize curved manifolds.  One of the fundamental problems in curved spaces  \cite{Fewster2008, Moretti2015} is that only for globally hyperbolic curved spacetimes endowed with a global temporal Killing vector,  the spacetime is a fiber bundle with a set of spatial slices or Cauchy surfaces which evolve in time.  For each fixed value of the temporal coordinate, one could solve the spectra of the Laplacian-Beltrami operator in the spatial slice as done for the Minkowski metric. However, in another more general case, if the curved spacetime is such that the fiber bundle does not allow an interpretation in terms of particle spectra independent of the observer, talking about scattering is ambiguous. Consequently, computing the quantum vacuum energy either from the $00$-component of the energy-momentum tensor and from the transfer operators defined in terms of the scattering data  will not offer a universal outcome independent of the observer. In fact, there are not many results about the \textit{TGTG}-formula in curved backgrounds,  and sometimes it does not even exist \cite{BordagJMMC2014}.  But there is a special case to be taken into account. When the frequencies of the particles created by the gravitational background are much smaller than the Planck frequency, one could use the perturbation theory for this curved spacetime as a semiclassical approach to quantum gravity \cite{Bar2009, Kay2006}.  In doing so, this weak gravitational backgrounds are treated classically and the matter fields are the ones which will be quantized.  The key point is that this gravity would be strong enough to produce some effects to the quantum matter, but not so strong as to require an own quantization. This is exactly the case which is going to be considered in this work.

Here,  I deal with the Casimir force between singular plates in a solitonic background.  In particular,  the plates are modeled by the Dirac point-like potential
\begin{eqnarray}
V_{\delta \delta'}(z)&=&v_0 \delta(z-a)+w_0 \delta'(z-a)+v_1 \delta(z-b)+w_1 \delta'(z-b),
\end{eqnarray}
being $v_0,v_1, w_0,w_1,a, b \in \mathbb{R}$ and $ a<b$. Above, $\delta'$ denotes the first derivative of the delta function.  Dirac delta potentials are widely used as toy models for realistic materials like quantum wires \cite{Cervero2002}, and to analyze physical phenomena such as Bose-Einstein condensation in periodic backgrounds \cite{BordagJPA2020},  or light propagation in 1D relativistic dielectric superlattices \cite{Halevi1999}. Despite being a rather simple idealization of the real system, the $\delta$ function has been proved to correctly represent surface interactions in many models related to the Casimir effect \cite{MiltonJPA}.  For instance, Dirac $\delta$ functions have been set on the plates acting as the electrostatic potential \cite{Hennig1992}, to represent two finite-width mirrors \cite{Fosco2009}, or to describe the permittivity and magnetic permeability in an electromagnetic context, by associating them to the plasma frequencies in Barton's model on spherical shells \cite{Barton2004, Parashar2012}.  On the other hand, the first derivative of the delta potential has been used to study monoatomically thin polarizable plates formed by lattices of dipoles \cite{Bordagprima2014},  and to analyze resonances in 1D oscillators \cite{Alvarez2013}. There is some controversy in the definition of the $\delta'$ potential since different regularizations produce different scattering data (see \cite{Toyama2007} and references therein). Here, I will use the one presented in \cite{Gadella2009}, in which the authors define it by introducing a Dirac delta potential at the same point to regularize the whole potential. As they explained, the major advantage of this choice is that it enables defining this singular potential in terms of matching conditions at the origin which do not depend on the choice of a regularization method.

In order to study the Casimir force between singular Dirac delta plates in a PT background,  the article is organised as follows: sections \ref{section3} and \ref{section4} involve the computation of the spectrum of scattering and bound states of the associated Schr\"odinger operator as well as the derivation of the Green's functions, respectively. In sections \ref{section5} and \ref{section6}, the \textit{TGTG}-formula and the DHN one \cite{DHN1974b} will be used to analyze the quantum vacuum interaction energy and to study the Casimir pressure between plates.  Finally,  section\ref{section7} summarizes the main conclusions. The natural system of units $\hbar=c=1$ will be used.

\section{Scattering data and spectrum}
\label{section3}
The Casimir force between plates is due to the coupling between the quantum vacuum fluctuations of the electromagnetic field with the charged current fluctuations of the plates \cite{Lifshitz1956}. For distances between plates rather larger than any other length scale concerning the electric response of the plates, only the long wavelength transverse modes of the electromagnetic field are relevant to the interaction. They can be mimicked by the normal modes of a scalar field whose dynamics is described by the action
\begin{eqnarray}
S[\phi] = \frac{1}{2} \int d^4x \left[\partial_\mu \phi \partial^\mu \phi -V_{\delta \delta'}(z)\phi^2-V_{PT} (z) \phi^2\right].
\end{eqnarray}

Consider for simplicity a real massless scalar field  $\phi$ confined between two parallel $(D-1)$-dimensional plates separated by a distance $b-a$ in the axis orthogonal to the plates, i.e. the $z$ axis.  For isotropic and homogeneous plates, there exists a translational symmetry along the surface of the plates,  and the theory of free fields without boundaries is recovered for the parallel direction coordinates $\vec{x_\parallel} \in \mathbb{R}^{D-1}$. Splitting the spatial coordinate as $x=(\vec{x_\parallel},z)$,  with $\vec{x_\parallel} \in \mathbb{R}^{D-1}$ and taking into account the Fourier decomposition of the field
\begin{equation}
\phi(t,x)= \int d\omega \, e^{-i\omega t} \phi_\omega(x),
\end{equation}
the equation for the modes of the fluctuations field is given by the non-relativistic Schr\"odinger separable eigenvalue problem
\begin{equation}\label{eqmodes}
(-\Delta + V_{PT}(z)+V_{\delta \delta'} (z))\phi_\omega(x)=\omega^2\phi_\omega(x).
\end{equation}
In the equation above,  $\Delta=\Delta_\parallel+\partial_z^2$,  and the frequencies follow the dispersion relation given by $\omega^2=\vec{k}_\parallel^{\, 2}+k^2$. Therefore, the problem is separable and only the direction orthogonal to the plates needs to be studied, as the spectrum is trivial in the other directions. From now on, I will consider $D=3$, just for simplicity. 

Notice that the non-relativistic Schr\"odinger operator related to the background in the dimension orthogonal to the plates, $\hat{K}_{PT}= -\partial_z^2 +V_{PT}(z),$ is not essentially self-adjoint in the Sobolev space of functions $W^2_2(\mathbb{R} -\{a,b\}, \mathbb{C})$.  It is necessary to add some matching conditions,  concerning the continuity of the wave function and the discontinuity of its derivative at the boundary points $\{a,b\}$, in order to define the self-adjoint extensions of $\hat{K}_{PT}$ in the aforementioned domain. These boundary conditions will be determined by the specific potential that represents the plates. For instance, if they are mimicked by the  Dirac delta potential $V_{\delta \delta'}$, the domain of the self-adjoint extension of $\hat{K}_{PT}$ is given by the suitable matching conditions:
\begin{eqnarray}\label{kurasov}
&&\mathcal{D}_{\hat{K}_{PT}} \!\!= \!\!\left\{ \phi \in W^2_2(\mathbb{R} -\{a,b\}, \mathbb{C}) \,  \Bigg| \,  \left(
\begin{array}{l}
\phi(a^+)\\
\phi^{\prime}(a^+)\\
\phi(b^+)\\
\phi^{\prime}(b^+)
\end{array}\right)=
\left(
\begin{array}{cccc}
\alpha_0 & 0 &0 &0\\[0.5ex]
\beta_0 & \alpha_0^{-1}&0 &0\\
 0 &0 & \alpha_1 & 0 \\[0.5ex]
0 &0 &  \beta_1 & \alpha_1^{-1}
\end{array}
\right)\left(
\begin{array}{l}
\phi(a^-)\\
\phi^{\prime}(a^-)\\
\phi(b^-)\\
\phi^{\prime}(b^-)
\end{array}\right)\right\}, \nonumber\\
&&\textrm{being} \qquad\alpha_{i}=\frac{1+w_i/2}{1-w_i/2}, \quad \beta_i= \frac{v_i}{1-(w_i/2)^2}, \qquad i=0,1,
\end{eqnarray}
coming from the original work of Kurasov in one-dimensional systems \cite{Gadella2009, Kurasov1996}.
The system of two plates in the background chosen has an open geometry so the positive energy spectrum will be continuous.  Scattering states correspond to solutions of the Schr\"odinger equation 
\begin{equation}
(\hat{K}_{PT}+V_{\delta \delta'}(z))\phi_{\omega, \vec{k}_\parallel}(z) = k^2 \phi_{\omega, \vec{k}_\parallel}(z),
\end{equation}
with $k\in\mathbb{R}$ (such that $k^2 > 0$),  and keeping in mind that the delta potential must be understood as the aforementioned boundary conditions at $z=a,b$.  Given a linear momentum $k$,  there are two independent scattering solutions to be found.  In order for the Casimir energy between plates to be a non-negligible magnitude, the two objects have to be very close to each other.  If the distance between plates is smaller than the support of the background potential in such a way that the plates are placed within this support,  the scattering solutions for particles incoming to the system from the left (also called right-handed solutions) and from right (left-handed solutions),  are respectively of the form:
\begin{eqnarray}
&&\hspace{-10pt}\psi^R_k(z)= \left\{ \begin{array}{lll}
              f_k(z)+r_R \, f_{-k}(z) , && \textrm{if} \quad z<a,\\[1ex]
              B_R\,  f_k(z) +C_R\, f_{-k}(z) , && \textrm{if} \quad a<z<b, \\[1ex]
              t_R \, f_k(z),  && \textrm{if} \quad z>b,
             \end{array}
   \right. \label{soldiestro}
   \end{eqnarray}
   \begin{eqnarray}
 &&\hspace{-10pt}\psi^L_k(z)= \left\{ \begin{array}{lll}
            t_L \, f_{-k}(z), && \textrm{if} \quad z<a,\\[1ex]
              B_L\,  f_{k}(z)+C_L\,\,  f_{-k}(z), && \textrm{if} \quad a<z<b, \\[1ex]
             r_L \, f_k(z) +  f_{-k}(z), && \textrm{if} \quad z>b.
             \end{array}
   \right. \label{solzurdo}
\end{eqnarray}
Notice that $f_k(z)=e^{ikz}(\tanh (z)-ik)$ are the eigenfunctions of the operator $-\partial_z^2+V_{PT}(z)$ (i.e. plane waves times first order Jacobi polynomials).  It is relevant to highlight that the transmission amplitudes $t_R(k), t_L(k)$ are identical to each other due to the time-reversal invariance of the Schr\"odinger operator. Consequently, they will be substituted by $t(k)$ from now on.  Replacing \eqref{soldiestro} and \eqref{solzurdo} in the matching conditions \eqref{kurasov} and solving the resulting two systems of equations with unknowns $\{r_R, r_L, t, B_R, B_L, C_R, C_L\} (k)$, the scattering data are obtained (they are collected in eq. \eqref{scatdatafull} in appendix \ref{AppendixA}).

The denominator of all the scattering parameters, $\Upsilon(k)$, is the spectral function (see its derivation in Appendix \ref{AppendixA}). The set of zeroes of $\Upsilon(k)$ can be the poles of the scattering matrix $S(k)$. Notice that the $S(k)$-matrix admits an analytic continuation to the entire complex momentum plane.  The zeroes of the spectral function on the positive imaginary axis in the complex momentum $k$-plane gives the bound states of the spectrum of the non-relativistic Schr\"odinger operator. Making $k \to i \kappa$ in $\Upsilon(k)=0$, one can study the bound states as the intersections between an exponential and a rational function  via  the transcendent equation
\begin{equation}\label{spec1}
-\frac{\Upsilon_1(\kappa)}{\Upsilon_2(\kappa)}= e^{-2\kappa(b-a)},
\end{equation} 
where
\begin{eqnarray}\label{spectralfull}
&&\Upsilon_1 (\kappa)\!=\! 4 \, \Sigma(v_0, w_0, a, \kappa) \, \Sigma(v_1, w_1, b, \kappa),\nonumber\\
&&\Upsilon_2 (\kappa) \!=\! 16 (\kappa-\tanh a)(\kappa+\tanh b) \, \Lambda(v_0, w_0, a, \kappa)  \Lambda(-v_1, w_1, -b, \kappa),\nonumber\\
&&\Lambda(v_i, w_i, x, \kappa)\!=\!-2  w_i \sech^2 x - (v_i -2 w_i \,\kappa)(\kappa-\tanh x),\nonumber\\
&&\Sigma(v_i, w_i, x, \kappa)\!=\! [2v_i +  \kappa (4+w_i^2)](\kappa^2-1)+ 2 \sech^2 x \, (v_i+2w_i \tanh x).
\end{eqnarray}  
Once the momenta of the bound states are determined, their energies can be computed as $E=(i\kappa)^2<0$. The lowest energy state will be characterized by $E_{min}$. For some type of potentials (such the pure delta plates, i.e. $w_0=w_1=0$), it is possible to give an analytic formula to bound the energy of the states with negative energy of the spectrum.  In the pure delta case,  this energy takes the value
\begin{eqnarray}\label{Emin}
\!\! \!\!  E_{min}=-\frac{1}{16}\left[(|v_1|+|v_0|)+\sqrt{(|v_1|+|v_0|)^2+16}\right]^2\! ,
\end{eqnarray}
for all $a,b,v_0,v_1 \neq 0$.  In other configurations, $E_{min}$ has to be obtained numerically.

Finding the \textit{TGTG}-formula implies focusing only in the scattering problem for one single plate. If one of the delta plates is removed (for instance $v_1=w_1=0$),  the spectral function of the reduced system is:
\begin{eqnarray}\label{polroots}
&&(4+w_0^2)\, \kappa^3+2v_0 \kappa^2 -(4+w_0^2)\,\kappa -2v_0 \tanh^2 a + 4w_0 \tanh a \sech^2 a =0.
\end{eqnarray}
By studying the asymptotic behavior of \eqref{polroots},  as well as its maxima and minima for different values of the parameters, it can be seen that there may be two cases: only one bound state or two bound  states. There is no zero mode because the state with wave vector $k=0$ does not constitute a pole of the $S(k)$-matrix.  
The scattering data for the reduced system can be obtained from equation \eqref{scatdatafull},  and they are given by:
\begin{eqnarray}\label{scatteringkink}
\hspace{-10pt}t^\ell&=& \frac{-\alpha_0 W}{\alpha_0 f_{-k}(a) \left(\alpha_0 f'_k(a)-\beta_0 f_k(a)\right)-f_k(a) f'_{-k}(a)}, \nonumber\\
\hspace{-10pt}r_R^\ell&=& \frac{-f_k(a) \left[\left(\alpha_0^2-1\right) f'_k(a)-\alpha_0 \beta_0 f_k(a)\right]}{\alpha_0 f_{-k}(a) \left(\alpha_0 f'_k(a)-\beta_0 f_k(a)\right)-f_k(a) f'_{-k}(a)},\\
\hspace{-10pt}r_L^\ell&=&\frac{-f_{-k}(a)\left[\left(\alpha_0^2-1\right) f'_{-k}(a)-\alpha_0 \beta_0 f_{-k}(a)\right]}{\alpha_0 f_{-k}(a) \left(\alpha_0 f'_k(a)-\beta_0 f_k(a)\right)-f_k(a) f'_{-k}(a)},\nonumber
\end{eqnarray}
where $\alpha_0,  \beta_0$ are defined in \eqref{kurasov}, and $W$ is the  Wronskian
\begin{equation}
W \equiv W[f_{k}(a) , f_{-k}(a)]  = -2 i k (k^2+1).
\end{equation}
From now on,  the superscript $\ell,r$ in the scattering data indicates which plate is being considered: $\ell$ for the plate placed on the left of the system and $r$ on the right.  The subscript $R,L$ refers to ``\textit{diestro}" (right-handed) and ``\textit{zurdo}" (left-handed) scattering.  Notice that when  $v_1=w_1=0$, the scattering data is such that $B_R^\ell=t^\ell, B_L^\ell=r_L^\ell, C_R^\ell=0, C_L^\ell=1$. Similarly, setting $v_0=w_0=0$ one obtains the reduced scattering data when the plate on the right is the only one present in the system. In this case, $\{t^r, r_R^r, r_L^r\}$ are given by \eqref{scatteringkink} but replacing $\alpha_0 \to \alpha_1, \beta_0 \to \beta_1, a \to b$. Furthermore, $C_L^r=t^r, B_L^r=0, B_R^r=1, C_R^r=r_R^r$.  

A rather important fact is that due to the P\"oschl-Teller background potential, the translational invariance of the system is broken. $V_{PT}(z)$ breaks the isotropy of the space,  and consequently if $f_k(z)$ is a eigenfunction of the non-relativistic Schr\"odinger operator $\hat{K}_{PT}$, then $f_k(z+a)$ with $ a\in \mathbb{R}-\{0\}$ will no longer be another.  This means that the scattering data explicitly depend on the position of the plates in a non-trivial way.

For computing the vacuum interaction energy in the corresponding QFT for the general case in which $v_0,v_1,w_0,w_1 \in \mathbb{R}-\{0\}$, the value of the energy for the lowest energy bound state of the quantum mechanical problem explained in this section is essential. Since the bound state with the lowest energy is characterized by $E_{min}$, the mass of the fluctuations in the theory will be balance with this value  of the energy  for making fluctuation absorption impossible. The unitarity of the QFT sets this lower bound for the mass of the quantum vacuum fluctuations, so that the total energy of the lowest energy state of the spectrum will be zero.   Thus,  the spectrum of the associated Quantum Field Theory will consist of a set of discrete states with energies within the gap $[0, |E_{min}|]$,  and a continuum of scattering states with energies above the threshold $E=|E_{min}|=m^2$. The number of discrete states will be determined by the value of the coefficients $\{v_0,v_1,w_0,w_1\}$, with a maximum of three being possible.\footnote{Although each $\delta\delta'$ potential can hold at most two bound states, since the two plates are very close together,  the whole system of two plates in the PT background acts as a well not deep and wide enough to accommodate four bound states, but three.}  It is important to highlight that the value of $E_{min}$ must be computed for the whole system of two objects plus the background potential. This is not contradictory to the fact that if the Casimir energy is calculated with the \textit{TGTG}-formula, only the transmission and reflection coefficients of the reduced system of an object in the background potential are needed. It will be discussed in detail in section \ref{section5}.

\section{Green's function}
\label{section4}
Once the spectral problem has been solved, the usual second quantization procedure  could be applied to promote the non-relativistic quantum mechanical theory to a QFT in which to study the quantum vacuum interaction energy between objects. The \textit{TGTG}-formula is based on two main elements: the Green's function,  and the transfer operators. The characteristic Green's function can be obtained by solving the differential equation
\begin{eqnarray}
\!\left[\partial_\mu \partial^\mu\! \!-2\sech^2 z+V_{\delta\delta'}(z)+m^2\right]G(x^\mu, y^\mu)=\delta(x^\mu-y^\mu)
\end{eqnarray}
for the complete Green's function 
\begin{equation}
G(x,x') = \int \frac{d^{2}k_\parallel}{(2\pi)^{2}} e^{i \vec{k}_\parallel (\vec{x}_\parallel-\vec{x'}_\parallel)} \int \frac{d\omega}{2\pi} e^{-i \omega (t-t')}G_k(z,z'),
\end{equation}
or equivalently by solving
\begin{eqnarray}
\! \left[-\partial^2_{z_1} \!-k^2\!-2\sech^2 z_1\!+V_{\delta\delta'}(z_1)\right]G_k(z_1,z_2)&=&\delta(z_1-z_2)
\end{eqnarray}
for the reduced Green's function.  Solving this differential equation requires assuming the continuity of  $G_k(z_1,z_2)$ and the discontinuity of its first derivative at the points $\{a,b\}$, as well as imposing an exponentially decaying behaviour of the solutions at infinity. Another way to compute the reduced Green's function in the spatial dimension orthogonal to the surfaces of the plates is by using \cite{Guilarte2015}
\begin{equation}
G_k(z,z') \!= \!\frac{\textrm{u}(z-z')\psi_k^R(z)\psi_k^L(z')\!+\! \textrm{u}(z'-z)\psi_k^R(z')\psi_k^L(z)}{W[\psi_k^R, \psi_k^L]} 
\end{equation}
with the two linear independent scattering solutions given in \eqref{soldiestro}-\eqref{solzurdo} for the complete system of two plates in the PT background. Note that $\textrm{u}(z-z')$ is the unit or Heaviside step function. Both aforementioned methods yield the same solution for the correlator.

Moreover, the Wronskian $W[\psi_k^R, \psi_k^L]$ has to be the same for the three zones in which the two delta plates divide the space. This imposes the following relation between the scattering coefficients: $t= B_R\,C_L-C_RB_L$. This relation is useful to simplify the solutions of the Green's function in the different zones, and to rewrite them as 
\begin{equation}G_k(z_1, z_2)=G^{PT}_k(z_1,z_2) +\Delta G_{k}(z_1,z_2),\end{equation} being $\Delta G_{k}(z_1,z_2)$ given by the equation \eqref{Green1} in appendix \ref{AppendixA}.

Notice that the Green's function for the kink potential centered at the origin without any delta interactions (i.e. $v_0=w_0=v_1=w_1=0$) takes the form
\begin{eqnarray}
&&\hspace{-18pt}G^{PT}_k (z_1, z_2)=\frac{1}{W} f_{-k}(z_<)f_{k}(z_>) =\frac{e^{ik|z_1-z_2|}}{W} \left(k^2+ik|\tanh z_1-\tanh z_2|+\tanh z_1 \tanh z_2\right),\nonumber\\&&
\end{eqnarray}
where $z_<$ and $z_>$ are the lesser or the greater of $z_1$ and $z_2$. It plays the same role as the Green's function  $G_k^0(z_1, z_2)=e^{ik|z_1-z_2|}/(-2ik)$ in free plain backgrounds. This is due to the fact that the P\"oschl-Teller potential is transparent (there is not additional reflection with respect to the free case). Furthermore, since the P\"oschl-Teller potential breaks the isotropy of the space, the Green's function is such that $G^{PT}_k (z_1,z_2) \neq G(z_1-z_2)$.  In fact, $G^{PT}(z,z)$ is not a constant as happens in the free flat case, but depends on the spatial orthogonal coordinate in a non-trivial way and thus, spatial translations are no longer symmetries of the system.

\section{Casimir energy and \textit{TGTG} formula}
\label{section5}

The quantum vacuum energy  per unit area of the plates is computed as the summation over all the frequencies $\omega=\sqrt{\vec{k}_\parallel^2+k^2+m^2}$ of the fields modes.  Notice that 
\begin{equation}
\omega^2 \in \sigma(-\partial^2_{\vec{x}_\parallel} + \hat{K}_{PT}+V_{\delta \delta'} (z)),
\end{equation}
where $\sigma$ refers to the spectrum of the operator.  However, 
\begin{eqnarray}
&&\frac{\tilde{E_0}}{A}=  \frac{1}{2} \,\, \underset{k}{\mathclap{\displaystyle\int}\mathclap{\textstyle\sum}} \,\,  \int_{\mathbb{R}^2} \frac{d\vec{k}_\parallel}{(2\pi)^2} \sqrt{m^2+k^2+\vec{k}_\parallel^2},
\end{eqnarray}
has a dominant ultraviolet divergence because the energy density of the free theory in the bulk gives an infinite result.  This is so because the bulk in this case involves the space between the plates,  and although the plates are separated by a finite small distance,  their surface area is intended to be infinitely long.  So the volume of the bulk is infinitely large.  The fact that the plates have infinite area also means that the self-energy of each plate is a divergence of lesser degree than that of the bulk. This divergence is often called subdominant divergence \cite{Asorey2013}.  The aim of this work is computing the quantum vacuum interaction energy,  which is the only part of the quantum vacuum energy $\tilde{E_0}$ that depends on the distance between plates,  and gives a finite result.  In order to obtain this quantity,  an exponentially decaying function acting as a regulator is introduced to remove the aforementioned divergences.  In such a way,  one rewrites the integration over the parallel modes as:
\begin{eqnarray}
&&\lim_{\epsilon \to 0} \int_{\mathbb{R}^2} \frac{d\vec{k}_\parallel}{(2\pi)^2}  \sqrt{\vec{k}_\parallel^2+k^2+m^2} \, e^{-\epsilon (\vec{k}_\parallel^2+k^2+m^2)}= \lim_{\epsilon \to 0} \frac{1}{2\pi}\, \chi(k,\epsilon) \, e^{-\epsilon (k^2+m^2)},
\end{eqnarray}
with
\begin{eqnarray}
 &&\chi(k,\epsilon)= \!\frac{\sqrt{\pi}}{4\, \epsilon^{3/2}}+\frac{\sqrt{\pi} (m^2+k^2)}{4 \sqrt{\epsilon}}-\frac{1}{3}\left(k^2+m^2\right)^{3/2}\!\!+o(\epsilon).
\end{eqnarray}
Eliminating the contribution of the parallel modes to the dominant and subdominant divergences means removing the first two terms in $\chi(k,\epsilon)$ before performing the limit.  In this way
\begin{eqnarray}\label{E01}
&&\frac{\tilde{E_0}}{A}=  -\frac{1}{2} \,\, \underset{k}{\mathclap{\displaystyle\int}\mathclap{\textstyle\sum}} \,\,  \frac{(m^2+k^2)^{3/2}}{6\pi}.
\end{eqnarray}
In the equation above, the sum over modes of the spectrum in the orthogonal direction splits into the summation over a finite number of states with positive energy in the gap $[0,|E_{min}|]$ (coming form the bound states of the associated quantum mechanical problem) and the integral over the continuous states with energies greater than $|E_{min}|$.

Now, it is necessary to remove in \eqref{E01} the contribution to the divergences coming from the modes in the orthogonal direction, where the one-dimensional kink lives. These contributions are different from the ones of the modes in the parallel directions, because in the orthogonal direction the space is no longer a free one. The usual method to be used is to put the system into a very large box of length $L$ with periodic boundary conditions (p.b.c.) at its edges
\begin{eqnarray}
&&\psi\left(-\frac{L}{2}\right)=\psi\left(\frac{L}{2}\right), \qquad \psi'\left(-\frac{L}{2}\right)=\psi'\left(\frac{L}{2}\right).
\end{eqnarray} 
By so doing, all the spectrum of the Schr\"odinger operator $\hat{K}=-\partial_z^2-2 \sech^2(z)+V_{\delta \delta'}(z)$ becomes discrete.  This fact will be taken into account below.

On the one hand, the contribution of the discrete set of $N$ states in the gap to the vacuum interaction energy is 
\begin{equation}
-\frac{1}{2}\sum_{j=1}^N\frac{\left(\sqrt{(i\kappa_j)^2+m^2}\right)^{3}}{6\pi}.
\end{equation} 
The frequencies of these bound states for each configuration $(v_0, v_1,w_0,w_1, a, b)$ will be determined numerically by solving $\Upsilon(i\kappa)=0, \,\kappa>0$ from the scattering problem. If there were half-bound states in the spectrum (i.e. states with energies that lie in the threshold $E=|E_{min}|$), they would have to be accounted for with a weight of $1/2$. But this will not be the case covered in the example of the two Dirac plates in a PT background.

\begin{figure}[t]
\centering
\includegraphics[width=0.45\textwidth]{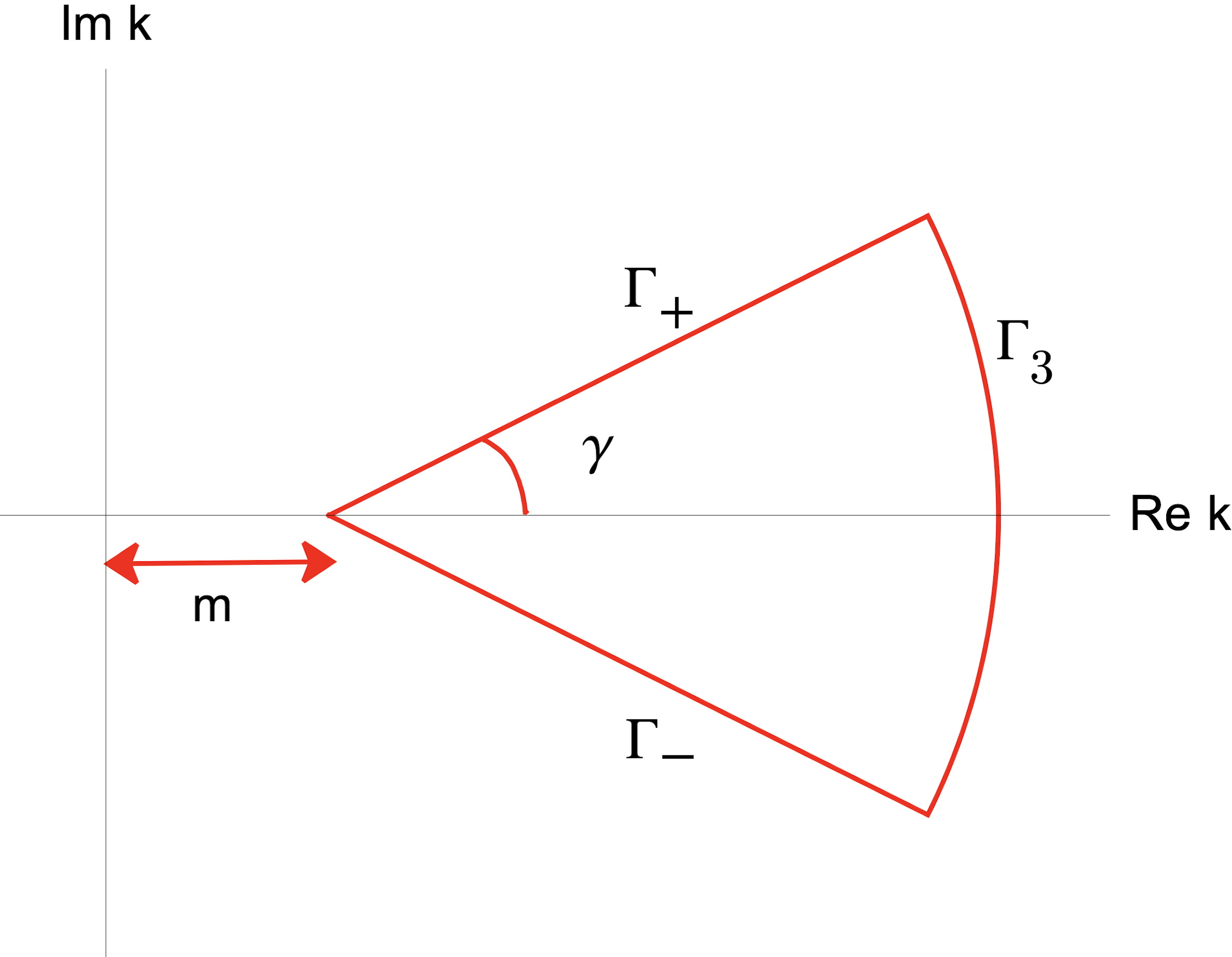}
\caption{\label{fig:fig1} Complex contour that encloses all the zeroes of the spectral function when $R \to \infty$. In this contour,  $\Gamma_\pm=\{m+\xi e^{\pm i\gamma}\, |\, \xi \in [0,R]\}$ and $\Gamma_3=\{m+R e^{i\nu}\, |\, \nu \in [-\gamma, \gamma]\}$. The angle $\gamma=\pi/2$ is going to be chosen. When integrating, the contour will be run counterclockwise. }
\end{figure}

On the other hand, concerning the states with energy $E>m^2$, it is necessary to compute
\begin{equation}\label{discreteE0}
-\frac{1}{2}\sum_{k_n \in \sigma^+(\hat{K})}\frac{\left(\sqrt{k_n^2+m^2}\right)^{3}}{6\pi},
\end{equation}
being $\sigma^+(\hat{K}) = \{k_n \in \sigma(\hat{K})\,  | \, k_n^2+m^2 >|E_{min}|\}$. The differential equation $\hat{K}\psi(z)=k_n^2 \psi(z)$ with periodic boundary conditions at $\pm L/2$ must be solved. Notice that now  the wave function $\psi(z)=A \psi_k^R(z) + B \psi_k^L(z)$ is a linear combination of the scattering solutions \eqref{soldiestro} and \eqref{solzurdo}. The resulting system of equations admits a solution whenever the following spectral equation holds:
\begin{eqnarray}
&&h_{p.b.c.}(k,L)\equiv 2t W -2 (t^2-r_R r_L) f_k\left(\frac{L}{2}\right)f_{k}^\prime\left(\frac{L}{2}\right)+2 f_{-k}\left(\frac{L}{2}\right)f_{-k}^\prime\left(\frac{L}{2}\right)\nonumber\\
&&+ (r_R+r_L)\left[f_{-k}\left(\frac{L}{2}\right)f_{k}^\prime\left(\frac{L}{2}\right)+f_{k}\left(\frac{L}{2}\right)f_{-k}^\prime\left(\frac{L}{2}\right)\right]=0.
\end{eqnarray}
The scattering data are again given in equation \eqref{scatdatafull} of appendix \ref{AppendixA}. The discrete set of zeroes $\{k_n\}$ of the secular function $h_{p.b.c.}(k,L)$ on the real axis will be the frequencies of the modes over which one has to perform the summation in \eqref{discreteE0}. This sum can be computed through a complex integral  over a contour enclosing all the zeroes of $h_{p.b.c.}(k,L)$. By using the residue theorem in complex analysis and also by taking into account the states in the gap, the total quantum vacuum interaction energy reads:
\begin{eqnarray}\label{E0DHN}
\frac{E_0}{A}&=& -\frac{1}{2} \left[\oint_\Gamma \frac{dk}{2\pi i} \frac{(m^2+k^2)^{3/2}}{6\pi}  \partial_k \log  h_{p.b.c.}(k,L) \right]-\frac{1}{2}\sum_{j=1}^N\frac{\left(\sqrt{(i\kappa_j)^2+m^2}\right)^{3}}{6\pi} ,
\end{eqnarray}
being $\Gamma$ the contour represented in FIG.  \ref{fig:fig1}.  It can be proved that the integration over the circumference arc of the contour is zero in the limit $R \to \infty$. Hence, the integration over the whole contour $\Gamma$ reduces to the integration over the straight lines  $\xi_\pm=\pm i \xi+m $ with $\xi\in [0, R]$.

Moreover, the dominant and subdominant divergent terms associated to the orthogonal modes and caused by the confinement of the system in a very large box must be subtracted as explained in \cite{Bordag2020, JMMC2020, Santamaria2019}, i.e. by computing:
\footnotesize
\begin{eqnarray}
\int_{0}^R \! \!\! \frac{d\xi}{12\pi^2 i}  \left\{\!(m^2+\xi_+^2)^{\frac{3}{2}} \! \left[L-L_0-\partial_\xi \log \frac{h_{p.b.c.}(\xi_+,L)}{h_{p.b.c.}(\xi_+,L_0)}\right]  -(m^2+\xi_-^2)^{\frac{3}{2}} \! \left[L-L_0-\partial_\xi \log \frac{h_{p.b.c.}(\xi_-,L)}{h_{p.b.c.}(\xi_-,L_0)}\right]\!\right\}
\end{eqnarray}
\normalsize
in the limits $L_0,R \to \infty$. The result of the integration does not depend on the box size and consequently, one could study the limit $L\to \infty$.  At this point, reversing the change of variable $\xi \to -ik$, due to the Wick rotation on the momentum, yields the DHN formula \cite{DHN1974b}
\begin{eqnarray}\label{E0old}
E_0=&& -\frac{A}{12\pi^2} \int_m^\infty dk \,   \left(\sqrt{k^2+m^2}\right)^3 \, \deruno{\delta(k)}{k}-\frac{A}{2}\sum_{j=1}^N\frac{\left(\sqrt{-(\kappa_j)^2+m^2}\right)^{3}}{6\pi},
\end{eqnarray}
with $m^2=|E_{min}|$,  and being  $\delta(k)$ the phase shift related to the scattering problem in the direction orthogonal  to the plate:
\begin{equation}
\delta(k)=\frac{1}{2i} \log_{-\pi} \left[t^2(k)-r_R(k)r_L(k)\right].
\end{equation}

So far, I have considered the scattering problem for waves interacting with a system inside a large box, without specifying the type of system I was working with. In addition, the divergences related to putting the system in a box were eliminated. However, notice that the system is composed by two infinitely large plates and a background potential. Consequently, the subdominant divergences associated to the plates are still present and a renormalization  mode-by-mode is necessary. This step is achieved by subtracting from the phase shift of the whole system with two plates, the corresponding phase shifts associated to a reduced problem with only one delta plate:
\begin{equation}\label{psmode}
\tilde{\delta} (k) = \delta_{v_0w_0,v_1w_1} (k)-\delta_{v_0w_0} (k)-\delta_{v_1w_1} (k).
\end{equation}
It is worth highlighting that this last equation constitutes a subtraction mode by mode of the spectrum to complete the renormalization. This method is different from the frequently used one of setting a cutoff in the integral over modes to remove the high energetic part of the spectrum that does not feel the background. The phase shift \eqref{psmode} has to be used in the DHN formula \eqref{E0old} to obtain a finite result.

Nevertheless, instead of using the aforementioned approach with the derivative of the phase shifts acting as the density of states, the \textit{TGTG}-formula will be used. The results will be the same using either of these two procedures. However, with the \textit{TGTG} representation we do not have to work with the scattering of the complete problem,  but with that of the reduced problem of a single object in the background. Hence, the numerical computational effort is much lower.

In the seminal paper \cite{KK2008}, O. Kenneth and I. Klich give the following formula for the quantum vacuum interaction energy between two compact bodies $1, 2$ in one spatial dimensional flat spacetime:
\begin{equation}\label{KK2008for}
E_0=-i \int_0 ^\infty \frac{d\omega}{2\pi} \tr \log\, (1-T_1 G_{12}T_2 G_{21}).
\end{equation}
Notice that the case the authors considered does not present bound states with negative energy in the spectrum of the corresponding Schr\"odinger operator in quantum mechanics. On the contrary, these type of states must be included in the case I am considering.  Furthermore, in the appendices B and C of \cite{KK2008} the authors prove that for any pair of disjoint finite bodies separated by a finite distance,  and any Green's function that is finite away from the diagonal, the $TGTG$-operator  is trace-class. The modulus of its eigenvalues is less than one and $\log (1-TGTG)$ is well defined. A similar reasoning can be followed here for the system of a pair of two-dimensional plates, that are assumed not to touch, in the background of a kink.  Thus, the only step left to be taken is to calculate the \textit{T}-operators for each one of the plates.

From the well-known Lippmann-Schwinger equation
\small
\begin{equation}
\Delta G_k(z_1,z_2)=  -\int dz_3\, dz_4\, G_k^{PT}(z_1,z_3) T_k(z_3,z_4)G_k^{PT}(z_4,z_2),
\end{equation}
\normalsize
and $\tilde{K}_{z_1} G_k^{PT}(z_1-z_2) = \delta(z_1-z_2),$ it is easy to see that 
\small
\begin{equation}
- \tilde{K}_{z_2} \, \tilde{K}_{z_1} \, \Delta G_k(z_1, z_2) \!= \!\int \!dz_3\,  dz_4 \, \delta(z_1-z_3) T_k(z_3, z_4) \delta(z_4-z_2), 
\end{equation}
\normalsize
where $\tilde{K}_{z}= \hat{K}_{PT}(z)-k^2$. Notice that in the above formula, $\Delta G_k(z_1, z_2)$  corresponds to the Green's function of only one plate in the P\"osch-Teller potential, i.e. the one given in \eqref{Green1}  with the coefficients of one of the plates equal to zero.  Due to the absolute values contained in $G_k^{PT}$, to obtain the transfer matrix $T_k (z_1, z_2)$ corresponding to one plate, the only non-trivial contribution comes from the case in which one point is on the left and the other one on the right of the plate. Hence, since in that case the Green's function is given by $\Delta G_k(z_1, z_2)=(t-1)\, G_k^{PT}(z_1, z_2)$, one needs to compute:
\begin{equation}
T_k (z_1, z_2) = - (t-1) \tilde{K}_{z_2} \, \tilde{K}_{z_1}  \, G^{PT}_k (z_1, z_2).
\end{equation}
In order to obtain the transfer matrix associated to the plate on the right\footnote{For computing  the transfer matrix of the left-hand side plate, one sets $v_1=w_1=0$ in the transmission coefficient \eqref{scatdatafull} involved in the Green's function \eqref{Green1} and considers the case in which the left plate is centered at $a=0$. },  one assumes the plate sitting at  the origin ($b=0$) for simplicity and hence, one of the coordinates $z_1, z_2$ will be greater than zero and the other less than zero. Taking into account that
\begin{eqnarray}
e^{ik|z_1-z_2|} &=& e^{ik (|z_1|+|z_2|)}, \nonumber\\
|\tanh z_1 - \tanh z_2| &=& \tanh |z_1| + \tanh |z_2|, \nonumber\\
\tanh z_1  \tanh z_2 &=& -\tanh |z_1| \tanh |z_2|,
\end{eqnarray}
(both in the cases $z_1<0, z_2>0$ and $z_1>0, z_2<0$), it is possible to rewrite the free Green's function in the background of the kink as
\begin{eqnarray}
&&G_k^{PT}(z_1, z_2) = -\frac{1}{W} f_k(|z_1|)f_k(|z_2|).
\end{eqnarray}
Because the Green's differential equation
\begin{equation}
\left(-\partial_{|z|}^2-k^2-2\sech^2 z\right) f_{k}(|z|) = 0
\end{equation}
holds, and using the formulas for the derivatives of functions depending on absolute values:
\begin{eqnarray}
\deruno{f_k(|z|)}{z}&=&f_k'(|z|)\,  \textrm{sign}\, z,\nonumber\\
 \der{f_k(|z|)}{z}{2}&=&f_k^{''}(|z|)+  f_k'(|z|) \, 2 \, \delta(z), 
\end{eqnarray} 
the transfer matrix for the right plate can be written as 
\begin{eqnarray}
&&\hspace{-10pt}T_k(z_1, z_2) = \frac{t-1}{W} 4 \, \delta(z_1)\, \delta(z_2)\, f'_k(|z_1|)f'_k(|z_2|)=-4 \, \delta(z_1)\, \delta(z_2)\, \Delta G(z_1,z_2) \, \frac{f'_k(|z_1|)}{f_k(|z_1|)}\, \frac{f'_k(|z_2|)}{f_k(|z_2|)}\nonumber\\
&&=\frac{|W|^2}{k^4} \delta(z_1) \delta(z_2)  \Delta G_k(z_1, z_2).
\end{eqnarray}
The Green's function for one delta plate is defined as the following piece-wise function:
\begin{empheq}[left= {\hspace{-6pt}\Delta G_{k}(z_1,z_2)\!=\! \empheqlbrace}]{align}
&\begin{aligned}
  & \frac{r_L}{W} f_{k}(z_1)f_{k}(z_2), & &\!\text{if $z_1, z_2 > b$,} \\
  &\frac{r_R}{W} f_{-k}(z_1)f_{-k}(z_2), & &\!\text{if $z_1, z_2 <b$},\\
  &\frac{t-1}{W} f_{k}(z_>)f_{-k}(z_<),  & &\!\text{otherwise},
\end{aligned}
\label{deltaGb}
\end{empheq}
being the scattering data given in \eqref{scatdatafull} but for the case $v_0=w_0=0$.  Consequently, 
\begin{eqnarray}
T_k(z_1, z_2) =- \frac{W^*}{k^2} \delta(z_1) \delta(z_2)  \left\{ \begin{array}{ll}
r_L (b=0), \\
r_R (b=0),\\
1-t(b=0).
\end{array}\right.
\end{eqnarray}
Asterisk means complex conjugate. The Lippmann-Schwinger operator is related to the scattering matrix by $S=1-i \delta(\omega-\omega')T$. This implies normalizing $T$ so that the factor $k^{-2}$ cancels out.

 When the delta potentials which mimics the plate is evaluated at another point different from the origin, the just-computed result for $T_k(z_1, z_2)$ is valid once after performing the translation $z_1\mapsto z_1-b$ and $z_2\mapsto z_2-b$. Notice that in the definition of $T$ at a point different to $z=0$, translation must be understood as replacing the scattering coefficients at $z=0$ contained in its definition by the ones at $z=b$. Due to the PT potential, the isotropy of the spacetime is broken.  Hence,   $r_{R,L}(b) \neq r_{R,L}(0) e^{ikb}$, so the general translation $z_1\mapsto z_1-b$ and $z_2\mapsto z_2-b$ aforementioned must not be interpreted in this usual sense.  The $T$ operator for the right-hand side plate is thus given by
\begin{eqnarray}\label{Toperator}
&&\!T(z_1, z_2) \!=\!- W^* \delta(z_1\! -b) \delta(z_2\! -b)\!\left\{\begin{array}{lc}
\! r_L(b),& \!\! \!\text{$z_1, z_2 \to b^+$\!,} \\
\! r_R(b), & \!\! \!\text{$z_1, z_2 \to b^-$}\!,\\
\! 1-t(b),  & \!\!\! \text{otherwise},
\end{array}\right.
\end{eqnarray}
and analogously for the plate located at $z=a$.

Notice that the $T$-matrix is the probability amplitude for a particle to interact with the potential but without propagation. Hence, in the system of two plates mimicked by point-like delta potentials in the background of a kink,  the definition of the $T$-operator must depend on $\Delta G$ evaluated at the point at which the delta potential is centered, as it is the case in \eqref{Toperator}. It could not depend on an arbitrary point of the spacetime for causality not to be violated. Notice that in more general cases, when the potential representing each object is not supported at a point but a compact  interval,   $T$ is local. Although in this case $T$ would depend on the points that constitute the support of the potential, it does not violate causality because it does not depend on arbitrary points.

By definition  $G^{PT}(z_1, z_2)=\langle \mathcal{T}\phi(z_1) \phi(z_2)\rangle$,  expression in which the time-ordering operator product has been considered. All eigenstates of $\hat{K}_{PT}$ with fixed energy $k^2$ can be described in terms of the orthonormal basis of left and right P\"oschl-Teller free waves.  By labeling $R=f_{ k}(z)$ and $L=f_{-k}(z)$, the Green's function or propagator can be written in this basis as
\begin{eqnarray}\label{GreenPTbasis}
&&G^{PT}(a,b)=\frac{1}{W} \ket{L(a)}\bra{L(b)}, \qquad 
G^{PT}(b,a)=\frac{1}{W} \ket{R(b)}\bra{R(a)},
\end{eqnarray} 
being $a<b$,  and the trace of the \textit{TGTG}-operator behaves as:
\begin{eqnarray}\label{TGTG0}
\hspace{-10pt}\ket{L(a)} \bra{L(b)} T^r \ket{R(b)}= r_L^\ell(v_0,w_0,a,k) \, r_R^r(v_1,w_1,b,k).
\end{eqnarray}
It has been taken into account that $T\ket{R}=\ket{L}$ and vice versa. So, it is clear that the \textit{TGTG}-formula will involve the reflection coefficients, which depend explicitly on the position of each plate. The above formula is the only product of the \textit{T}-matrix components that allows coincidences of the $z_1, z_2$ points in $[a,b]$ and contributes to the quantum vacuum interaction energy between plates. In the special case discussed here, the $TGTG$ operator has rank equal to one. This implies that
\begin{eqnarray}\label{TGTG1}
\tr \log (1-TGTG) =\log \, \det(1-TGTG)= \log(1-\tr \, TGTG).
\end{eqnarray}
For other cases, if the modulus of the eigenvalues of the \textit{TGTG} operator is less than one, it is still possible to use 
\begin{eqnarray}\label{TGTGAPROX}
\tr \log (1-TGTG) &=&\log \, \det(1-TGTG)\approx  \log(1-\tr \, TGTG)
\end{eqnarray}
as a good approximation up to first order to simplify \eqref{KK2008for}. The demonstration is collected in appendix \ref{AppendixB}. In summary, replacing \eqref{TGTG0} and \eqref{TGTG1} in \eqref{KK2008for},  and generalizing it to three dimensions,  leads the final expression \eqref{finalTGTG} with $m^2=|E_{min}|$.  
\begin{eqnarray}\label{finalTGTG}
\!\! \frac{E_0}{A}&=&-\frac{1}{2}\! \sum_{j=1}^N\! \frac{\left(\sqrt{-\kappa_j^2+m^2}\right)^{3}}{6\pi} + \!\frac{1}{8\pi^2}\!\! \int_{m}^{\infty} \!\! \! \!\! d\xi\,  \xi\,  \sqrt{\xi^2-m^2} \, \textrm{log} (1-\textrm{Tr}\, TGTG_\xi),\\
&=&-\frac{1}{2}\!\sum_{j=1}^N\!\frac{\left(\sqrt{-\kappa_j^2+m^2}\right)^{3}}{6\pi}+\!\frac{1}{8\pi^2} \!\!\int_{m}^{\infty} \!\! \! \!\! \!\! d\xi\,  \xi\,  \sqrt{\xi^2-m^2} \, \textrm{log} \! \left[1\!-\!    r_L^\ell(v_0,w_0,a,i\xi) \, r_R^r(v_1,w_1,b,i\xi) \right]\!.\nonumber
\end{eqnarray}

There are some details that are worth highlighting. Firstly, if $f_\pm(z)$ were replaced by the usual plane waves, Kenneth and Klich's original \textit{TGTG}-formula would be restored. The reason is that,  in flat isotropic spacetimes,  the scattering coefficients for plates placed at another point different from the origin are equal to the ones at $z=0$ times an exponential factor that accounts for the translation that has taken place:
\begin{eqnarray}
&&r_L^\ell(a) = r^\ell_L(0) \, e^{-2iak} = r^\ell_L(0)\, W \, G_{-k}^{(0)}(-a,a) , \nonumber\\
&& r_R^r(b) = r_R^r(0) \, e^{2ibk} =r_R^r(0)\, W \, G_{k}^{(0)}(-b,b).
\end{eqnarray}
 On the contrary, the main difference when working with weak and transparent curved spacetimes that breaks the isotropy of the space is that this rule no longer applies. Thus, the scattering coefficients for plates placed at another point different form the origin are equal to the ones at $z=0$ times the quotient between the transmitted probability amplitude at the generic point and the one at $z=0$, and multiplied by another function related to the configuration of the space:
 \begin{eqnarray}
&&r_L^\ell(a) = r^\ell_L(0) \, \frac{ G_{-k}^{PT}(-a,a)  \, t(a)\, h^+(\alpha_0, \beta_0,-a)}{G_{-k}^{PT}(0,0)  \, t(0)\, h^+(\alpha_0, \beta_0,0)} , \nonumber\\
&&r_R^r(b) = r_R^r(0) \, \frac{G_{k}^{PT}(-b,b) \, t(b) \, h^-(\alpha_1, \beta_1,b)}{G_{k}^{PT}(0,0)  \, t(0) \, h^-(\alpha_1, \beta_1,0)} ,
\end{eqnarray}
being $h^\pm(\alpha_i, \beta_i,x)=\alpha_i \beta_i \pm  (\alpha_i^2-1)f'(k,x)/f(k,x)$. This fact is crucial to generalize the \textit{TGTG}-formula in the case studied. 

Secondly, by defining the potential  $V_i(z)$ to describe each of the two plates as 
\begin{equation}
V_i(z)=v_i \, \delta(z-z_i)+w_i \delta'(z-z_i)
\end{equation} (with $i=0,1$ and $z_0=a, z_1=b$),  the Schr\"odinger operators $\hat{K}_i=-\partial_z^2-V_{i}(z)$ are defined over a Hilbert space that, in general, is not isomorphic to that of $\hat{K}_{PT}$. Hence, $G^{PT}$ and $T^{r, \ell}$ do not act in the same spaces,  and $T^\ell G^{PT} T^r G^{PT}$ is ill-defined. To avoid this problem, a Wick rotation of the momentum $k$ must be performed in order for all the operators to act in the same Hilbert space.  The integral \eqref{finalTGTG} is thus  convergent and can be evaluated numerically with \textit{Mathematica}. In the next section, the results of the Casimir energy for some configurations of the plates in the PT background potential are going to be discussed.

\section{Casimir pressure}
\label{section6}

Once the quantum vacuum interaction energy is determined, one can study the Casimir force between plates as $F= -\,  \partial E_0/ \partial d$,  being $d$ the distance between plates. Nevertheless, as already explained, the translational invariance is broken due to the PT background, which means that the scattering data for the plates explicitly depend on the position in a non-trivial way.  Hence, when computing the Casimir force, a non-trivial contribution coming from the derivatives of the scattering amplitudes of one of the plates with respect to the position will appear. There is an ambiguity not yet  clarified in this calculation. One can either introduce the dependence on the distance between plates in three different ways:
\begin{enumerate}
\item Putting the left-hand side plate at $z_1=a$ and the right-hand side one at $z_2=a+d$. In this case, only the scattering data of the plate on the right will depend on the distance $d$, and only the derivative of $r_R^r(v_1,w_1,a+d,k)$ with respect to $d$ will appear.
\item Considering the right-hand side plate placed at $z_2=b$ and the left-hand side one at $z_1=b-d$. Analogously to the previous case, the derivative of $r_L^\ell(v_0,w_0,b-d,k)$ with respect to the distance is the only possible contribution.
\item When one of the plates is to the left of the origin and the other one to the right, one could describe the location of the plates as the left one being at $z_1=-d+b$ and the other one at $z_2=d+a$, with $a<0$ and $b,d>0$. This case is different because  the derivatives of the reflection coefficients of both plates $r_R^r(v_1,d+a ,k)$ and $r_L^\ell(v_0,-d+b,k)$  will be taken into account.
\end{enumerate}
It is work in progress to check that these three situations give rise to the same force. However, it seems reasonable to think that if the Casimir energy between plates has a change in the sign for some values of the parameters $\{v_0,v_1,w_0,w_1, a,b\}$, the Casimir force will present it too. Consequently, studying numerical results for the quantum vacuum interaction energy is enough to discuss whether this flip of sign appears as a consequence of the introduction of the $\delta'$ potential, as was the case in other configurations in flat spacetimes \cite{Guilarte2015, Romaniega2021}.

\begin{figure}[t]
\centering
\includegraphics[width=0.48\textwidth]{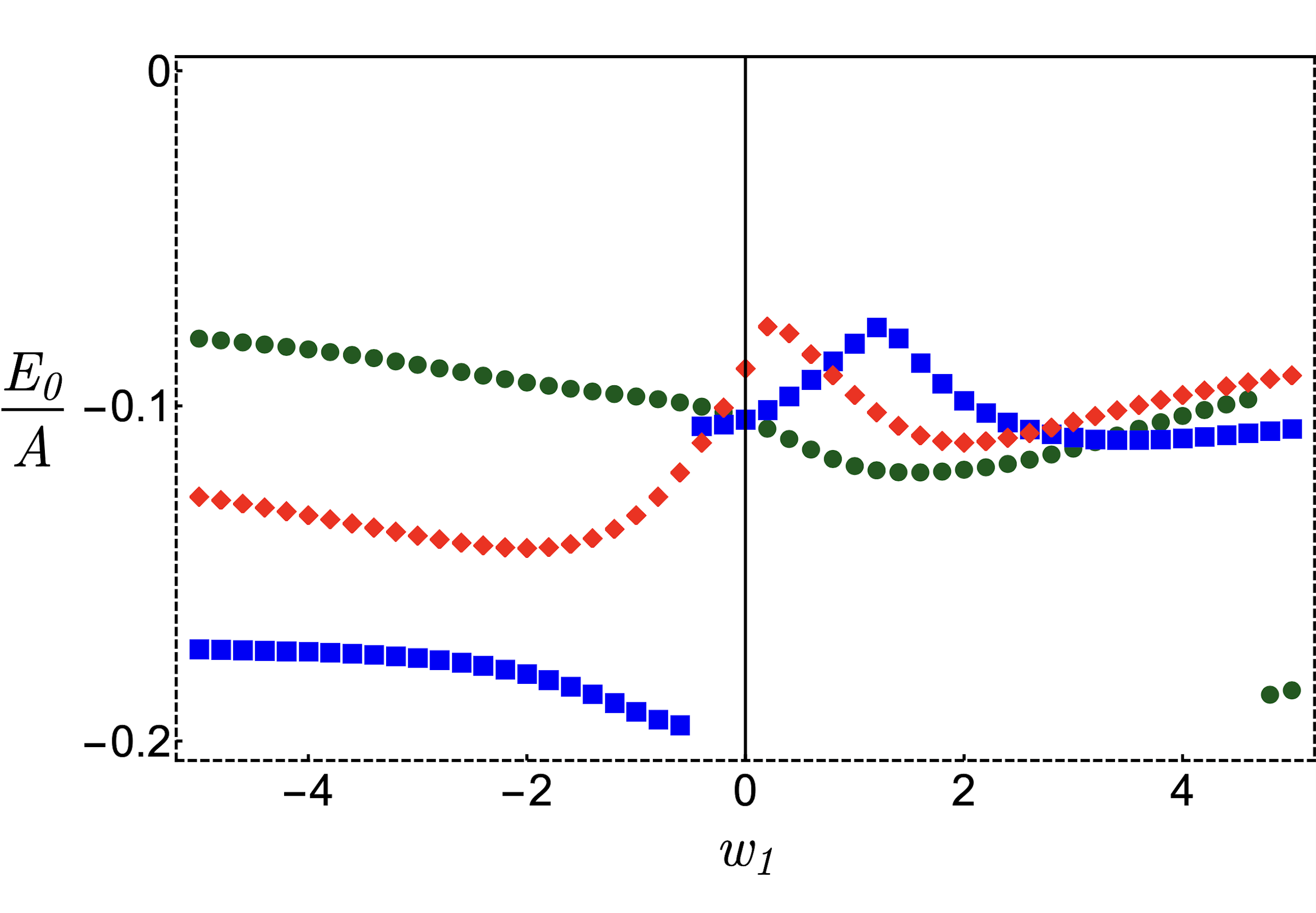}
\caption{Casimir energy per unit area between plates situated at $a=-0.2, b=0.8$, as a function of the coefficient $w_1$. Different configurations are shown: a) Pure $\delta'$ plates (i.e. $v_0=v_1=0)$ with $w_0=3$ (rhombi), b) Identical plates characterized by $v_0=v_1=1$ and $w_0=w_1$ (circles), c) Opposite plates described by $v_0=v_1=1$ and $w_0=-w_1$ (squares).  }
\label{fig:PTdentro0}
\end{figure}

\begin{figure}[t]
\centering
\includegraphics[width=0.48\textwidth]{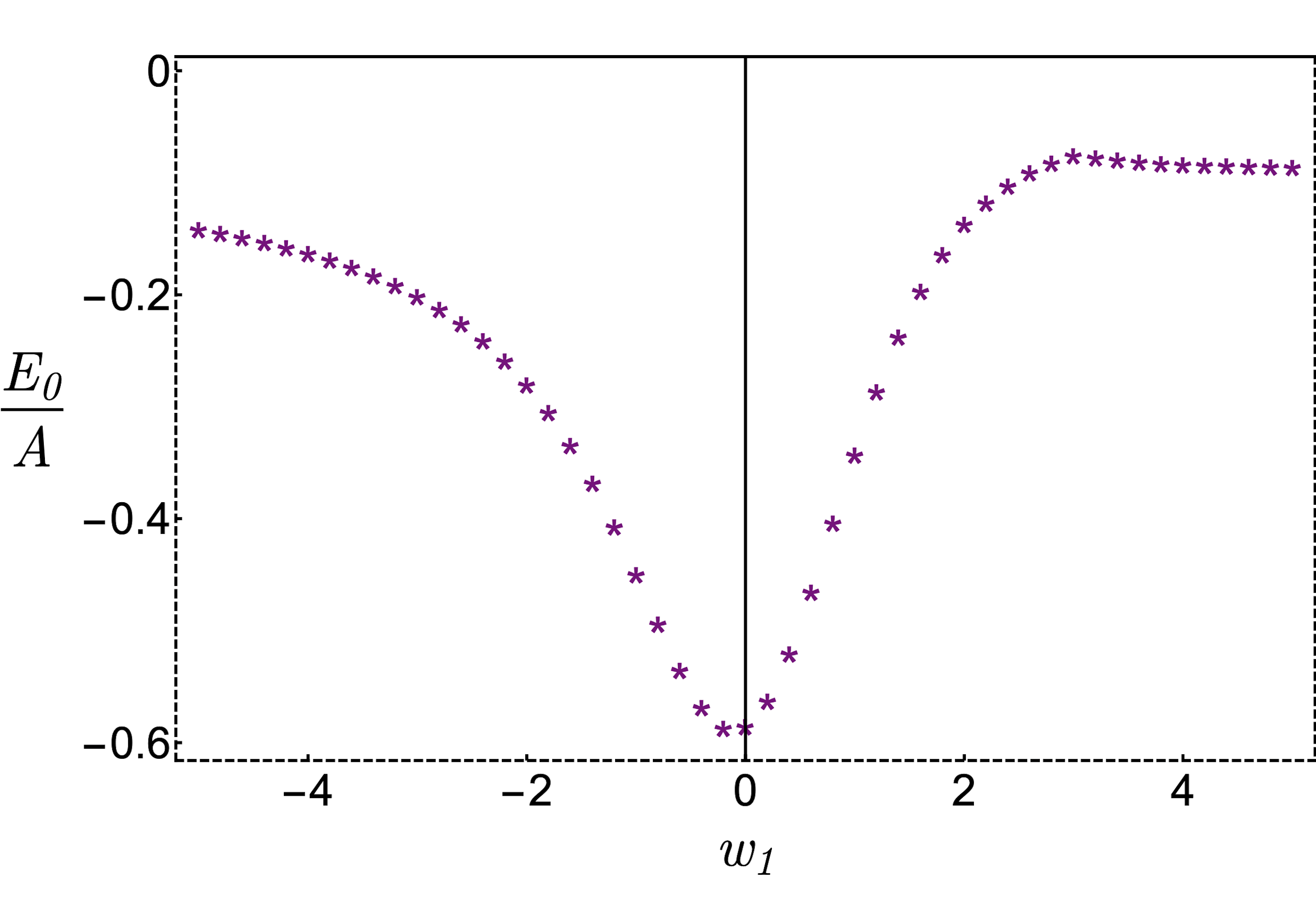}
\caption{Casimir energy  per unit area between plates situated at $a=-0.2, b=0.8$, as a function of the coefficient $w_1$. A generic $\delta\delta'$ potential with $v_0=1, v_1=-4, w_0=2.5$ has been considered.}
\label{fig:PTdentro1}
\end{figure}

\begin{figure}[t]
\centering
 \includegraphics[width=0.49\textwidth]{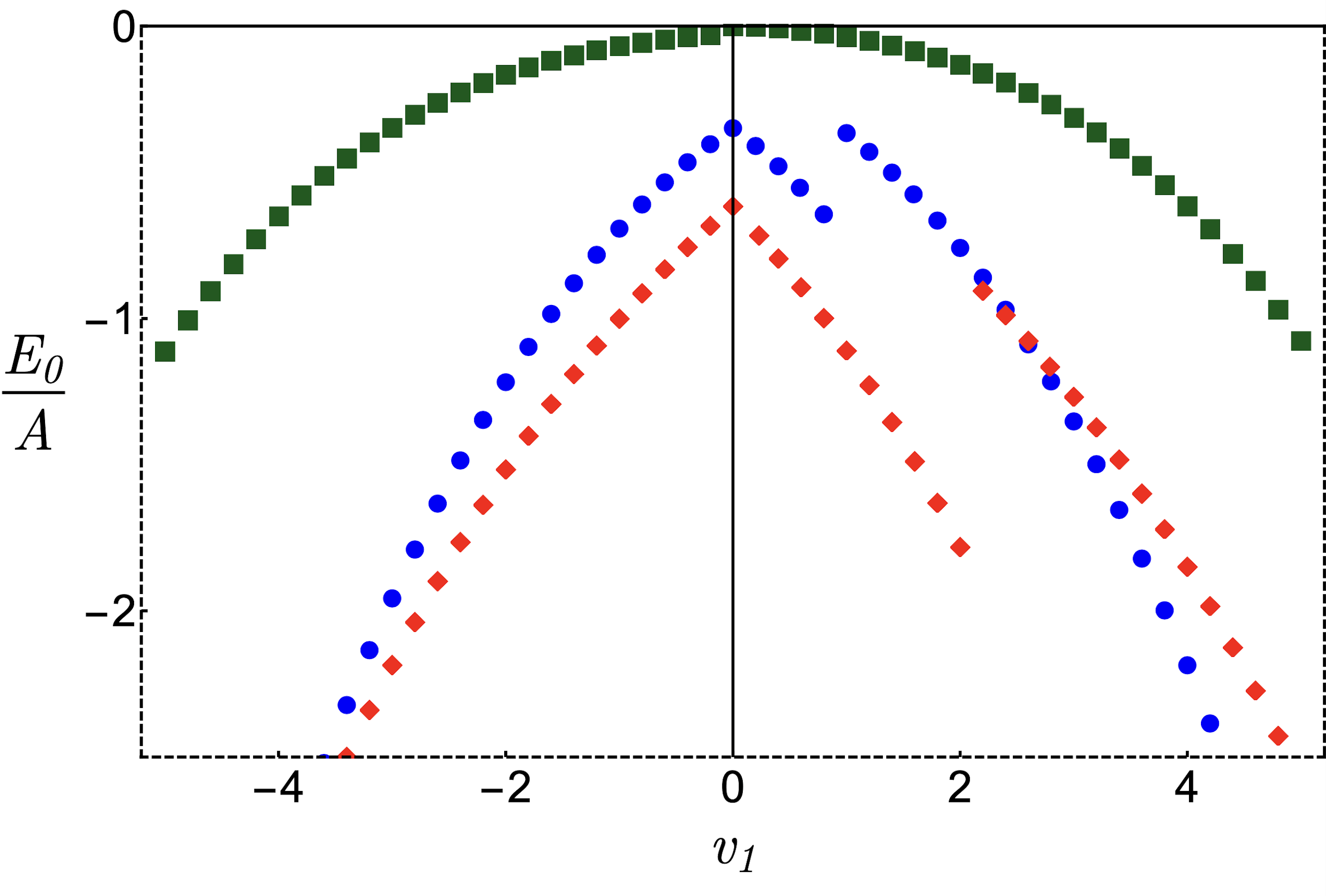}
\caption{Casimir energy  per unit area between pure delta plates ($w_0=w_1=0$) situated at $a=-b=-0.5$ as a function of $v_1$. In this plot $v_0=-3$ (circles), $v_0=0.1$ (squares) and $v_0=4$ (rhombi).}
\label{fig:PTdentro2}
\end{figure}

FIG.  \ref{fig:PTdentro0},  \ref{fig:PTdentro1} and \ref{fig:PTdentro2} show the quantum vacuum interaction energy per unit area of the plates for different configurations of the system of two plates in the PT background.
As can be seen in the figures above,  the energy is always negative,  independently on the value of the $\delta$ and $\delta'$ coefficients. Furthermore it could be checked numerically that the quantum vacuum interaction energy is definite negative,  regardless of the relative position of the plates with respect to the kink center as well. This implies that the Casimir force between plates will always be attractive in this system.  However, due to the changes of the spectrum of bound states as a function of $\{v_0,v_1,w_0,w_1\}$,  and the relative position between the plates and the kink, there is a peculiar fact characteristic of the spectra shown in FIG. \ref{fig:PTdentro0} and \ref{fig:PTdentro2}.  The sudden discontinuities present in the plots are related to the `loss' of one bound state with very low $k=i\kappa$ (nearly zero) in the spectrum of the system.  At this point, it is necessary to realize that the whole system of the two plates together with the PT potential acts as a well with a fixed depth and width.  Consequently, at the configuration in the space of parameters at which the jump appears, the resulting well is not deep enough to hold more bound states with large negative energy.  This loss of a bound state, which becomes a state in the continuum,  translates into a jump in the energy,  because the terms related to the states within the gap gives the major contribution to the total energy.   The jump discontinuities in the energy appear for the $\delta\delta'$ plates case whenever $v_i>0$. It can be checked that for the cases of identical and opposite plates, if $v_i<0$, the results would be qualitatively similar to those shown in FIG.  \ref{fig:PTdentro1}. Consequently, the introduction of the $\delta'$ potential modifies the spectrum in such a way that the well represented by the system can better accommodate bound states if $v_i<0$. In these cases, for different configurations of the coefficients of the $\delta\delta'$ functions very close to each other, no bound states with momenta close to zero are lost.

Another important conclusion can be drawn.  For flat spacetimes, it has been proved that the introduction of the $\delta'$ potential causes the sign of the force to change in different areas of the parameter space. This behaviour has been observed for instance for a scalar field and two concentric spheres defined by such a singular $\delta \delta^\prime$ potential on their surfaces \cite{Romaniega2021},   or in \cite{Guilarte2013} for $\delta \delta^\prime$ plates. However, curiously, when considering this last configuration in a sine-Gordon kink spacetime, the change of sign in the energy disappears.  Namely,  one can consider a curved spacetime that is confining (in the sense that the background acts as a well) and the plates are within the support of that well. Then,  even in the case where the plates act as very repulsive barriers, there are still negative energy states in the spectrum of the associated Schr\"odinger operator,  and the Casimir energy between plates will be attractive.  Consequently, although the background potential under consideration constitutes an example of weak curved background, the results are quite different from the flat  case. 

When the P\"oschl-Teller potential is not confined at all between the plates and they are far from the kink centre, the system of two $\delta\delta'$-plates in flat spacetime is recovered. This is the reason that the numerical representation for this situation, which has already been studied in the literature \cite{Guilarte2013},  is not included here. 

Finally, is worth pointing that even in the case where there is only one plate in the system, the other plate feels the interaction because there is still a non-zero quantum vacuum interaction energy in the system. This can be checked by looking at the non-zero values of the energy appearing in the vertical axis for the pure $\delta'$ plates (in this axis, $w_1=0$ and there is no right plate in the system) in FIG.  \ref{fig:PTdentro0}. 

\section{Conclusions}
\label{section7}
In this work, a quantum scalar field between two parallel two-dimensional plates in a sine-Gordon background at zero temperature is presented. The main result is the generalization of the \textit{TGTG}-formula for weak and transparent gravitational backgrounds in which the frequencies of the particles created by the gravitational background are much smaller than the Planck frequency, and the fields could be asymptotically interpreted as particles. The quantum vacuum interaction energy has been calculated using this formula, which only depends on the reflection coefficients associated to the scattering problem. They involve a dependence on the analogous of the plane waves in flat spacetimes but for the specific  background potential chosen. The Casimir energy thus depends on the parameters describing the potentials and on the distance from the plates to the center of the kink. For obtaining the energy, it has been necessary to compute the Green's functions from the scattering data. The transfer matrix has been determined with complete generality too, in terms of the Green's function.  Although the \textit{TGTG}-formula has the advantage that it depends only on the scattering data of one of the plates,  and it is not necessary to solve the scattering problem of the whole system, the well-known DHN formula has also been derived for completeness.  The virtue of the \textit{TGTG}-formula obtained here is that it can be easily generalized to other type of configurations, either for another background and for other potentials that could properly mimic the plates. 

As an example, two plates mimicked by Dirac $\delta$-potentials and its first derivative in a  background of a topological P\"oschl-Teller kink is studied. The quantum vacuum fluctuations around the kink solution could be interpreted as mesons propagating in the spacetime of a domain wall. One of the relevant characteristics of the  P\"oschl-Teller background is that the translational symmetry is broken and the space is anisotropic. This translates into the fact that the scattering coefficients, as well as the Green's function, will depend on the position of the plates in a non-trivial way.  The wave functions of the continuous spectrum of states with positive energy have been characterized by means of these scattering data.  The bound states have also been studied, setting a threshold for the minimum negative energy in the system. The unitarity of the QFT requires this lower bound be fixed as the mass of the quantum vacuum fluctuations,  so that the total energy of the lowest energy state of the spectrum will be zero, making fluctuations absorption impossible. It is worth highlighting that the quantum vacuum energy for this (3+1)-dimensional problem is only negative, independently on the value of the coefficients of the $\delta \delta'$ potentials and its location in relation to the kink center. This implies that the Casimir force between plates will be attractive in this system. Furthermore, even in the case where there is only one plate in the system, the other plate feels the Casimir interaction because there is still a non-zero quantum vacuum interaction energy in the system.   

Once the Casimir energy between plates in the background of a sine-Gordon kink has been computed by using the \textit{TGTG}-formula, it would be enlightening to obtain the same result but from the integration of the $00$-component of the energy-momentum tensor. In this way one could also study the spatial distribution of the energy density.  It is left for further investigations. 

I would like to finish by mentioning a possible future phenomenological application of the study collected in this work. Notice that in the Casimir effect, the electromagnetic force results from the computation of the one-loop quantum correction to the vacuum polarization in QED. Nevertheless, taking a quadratic approximation for the action implies that it does not depend on the coupling constant of the theory, and for that reason the force is relevant at the nanometre scale. By the same argument, if one-loop quantum corrections to the graviton propagator were calculated, the coupling constant of the gravitational theory would not appear either. And therefore, since the graviton has zero mass, one could think of studying the quantum interacting force between gravitational objects  separated by a small distance, using the same quadratic approximation reasoning applied here. The force caused by non-massive quantum vacuum fluctuations around a classical solution for gravitons coupled to no matter which other massive field would constitute an example of quantum corrections to the gravitational field theory. The difficulty will lie in finding a material that is opaque to the gravitational waves \cite{Quach2015}. Although interesting, this open problem is left for future research.

\section*{Acknowledgment}
I am grateful to M. Santander, C. Romaniega and L.M. Nieto for fruitful discussions. This research was supported by Spanish MCIN with funding from European Union NextGenerationEU (PRTRC17.I1) and Consejer\'ia de Educaci\'on from JCyL through QCAYLE project, as well as MCIN project PID2020-113406GB-I00.  I am thankful to the Spanish Government for the FPU PhD fellowship program (Grant No. FPU18/00957). 


\bibliographystyle{ptephy}
\bibliography{biblio}


\appendix

\section{Scattering data and Green's function for two $\delta\delta'$ plates in a PT background}
\label{AppendixA}

The scattering data as well as the spectral function are obtained directly when the system of equations \eqref{soldiestro},  \eqref{solzurdo} and \eqref{kurasov} is solved with the software \textit{Mathematica}.  More specifically, for the left-to-right case, if one replaces \eqref{soldiestro} in the matching conditions \eqref{kurasov} one obtains the following system:
\small
\begin{eqnarray}
[B_R f_k(z)+C_R f_{-k}(z)]\left.\right|_{z\to a} &=& \alpha_0 [f_k(z)+r_R f_{-k}(z)]\left.\right|_{z\to a}\nonumber \\
\frac{\partial}{\partial z}[B_R f_k(z)+C_R f_{-k}(z)]\left.\right|_{z\to a} &=& \beta_0 [f_k(z)+r_R f_{-k}(z)]\left.\right|_{z\to a} +\frac{1}{\alpha_0}\frac{\partial}{\partial z}[f_k(z)+r_R f_{-k}(z)]\left.\right|_{z\to a}\nonumber \\
t_R f_k(z)\left.\right|_{z\to b} &=& \alpha_1 [B_R f_k(z)+C_R f_{-k}(z)]\left.\right|_{z\to b}\nonumber  \\
\frac{\partial}{\partial z}[t_R f_k(z)]\left.\right|_{z\to b} &=& \beta_1 [B_R f_k(z)+C_R f_{-k}(z)]\left.\right|_{z\to b} +\frac{1}{\alpha_1}\frac{\partial}{\partial z}[B_R f_k(z)+C_R f_{-k}(z)]\left.\right|_{z\to b}\nonumber  \\&&
\end{eqnarray}
\normalsize
Solving this system of equations, the scattering data for the non-relativistic mechanical problem of scalar fields propagating in the curved background of a topological PT kink while interacting with two Dirac $\delta \delta'$ plates are obtained. They are given by \eqref{scatdatafull}.  The right-to-left scattering can be obtain analogously by replacing \eqref{solzurdo} in \eqref{kurasov},  and it is easy to check that the denominator of the scattering data is the same as in the left-to-right case.
\footnotesize
\begin{eqnarray}\label{scatdatafull}
t(k)&=&\frac{1}{\Upsilon(k)}\alpha_0 \alpha_1 W^2,\nonumber\\
r_R(k)&=&\frac{1}{\Upsilon(k)}[-f_k(b) f_k(a)\left(f'_{-k}(b) \left(f'_k(a)-\alpha_0 \mathcal{A}_0 (a)\right)+\alpha_0^2 f'_{-k}(a) \left(-\alpha_1 \mathcal{A}_1(b)+f'_k(b)\right)\right)\nonumber\\
 &&+\alpha_1 \mathcal{A}_1(b) f_k(a) f_{-k}(b) \left(f'_k(a)-\alpha_0 \mathcal{A}_0(a)\right)-f_k(b)f_{-k}(a) \left(f'_k(a)+\alpha_0 \beta_0 f_k(a)\right) \left(\alpha_1 \mathcal{A}_1(b)-f'_k(b)\right)],\nonumber\\
r_L(k)&=&-\frac{1}{\Upsilon(k)}[f_{-k}(b)f_k(a)f'_k(a)(-\alpha_1 \mathcal{A}_1^*(b)+f_{-k}'(b))+\alpha_0 \beta_0 f_{-k}^2(a)(-f_{-k}(b)f_k'(b)+\alpha_1 f_k(b)\mathcal{A}_1^*(b))\nonumber\\
&&+f_{-k}(a)[f_{-k}(b)\alpha_0 \mathcal{A}_0(a) (\alpha_1 \mathcal{A}_1^*(b)-f_{-k}'(b))+(-1+\alpha_0^2)f_{-k}'(a)(-\alpha_1 f_k(b)\mathcal{A}_1^*(b)+f_k'(b)f_{-k}(b))]],\nonumber\\
B_R(k)&=&\frac{1}{\Upsilon(k)}[\alpha_0 W \left(f_k(b) f'_{-k}(b)-\alpha_1 \mathcal{A}_1(b) f_{-k}(b)\right)],\nonumber\\
B_L(k)&=&-\frac{1}{\Upsilon(k)}[\alpha_1 W f_{-k}(a) \left(f'_{-k}(a)-\alpha_0 \mathcal{A}_0^*(a)\right)],\nonumber\\
C_R(k)&=&-\frac{1}{\Upsilon(k)}[\alpha_0 W f_k(b) \left(f'_k(b)-\alpha_1 \mathcal{A}_1(b)\right)],\nonumber\\
C_L(k)&=&-\frac{1}{\Upsilon(k)}[\alpha_1 W \left(\alpha_0 \mathcal{A}_0(a) f_{-k}(a)-f_k(a) f'_{-k}(a)\right)],\nonumber\\
\Upsilon(k)&=&-\left(\alpha_0 \mathcal{A}_0^*(a)-f'_{-k}(a)\right) \left[f_k(b) \left(f_{-k}(a) \left(\alpha_1 \mathcal{A}_1(b)-f'_k(b)\right)+f_k(a) f'_{-k}(b)\right)-\alpha_1 \mathcal{A}_1(b) f_k(a) f_{-k}(b)\right]\nonumber\\
&&+\alpha_0^2 \, W \left(f_k(b) f'_{-k}(b)-\alpha_1 \mathcal{A}_1(b) f_{-k}(b)\right).
\end{eqnarray}
\normalsize
The notations $\mathcal{A}_i(z)=-\beta_i \, f_k(z)+\alpha_i \, f'_k(z)$,  together with  $W=-2ik(k^2+1)$,  have been used to simplify the expressions.  

Notice that the denominator of all the scattering parameters, $\Upsilon(k)$, is the Jost function.  The spectral or Jost function is widely used in the literature  \cite{Taylorbook, Galindobook, Jost1952a, Jost1952b} because of its connection to the denominator of the scattering coefficients.  Notice that the Jost function $j(k)$ is related to the $S$-matrix by means of $\det S = t^2(k)-r_R(k) r_L(k)= e^{i 2 \delta(k)}=j^*(k)/j(k)$.  It should be highlighted that the phase shift in the scattering problem is just minus the phase of the Jost function.  Furthermore, the zeroes of the Jost function on the positive imaginary axis of the complex momentum plane characterize the wave vector of the bound states in the quantum mechanical problem.  Consequently, the Jost function determines both the spectrum and the phase shift of the scattering problem.   From the last expression of $\Upsilon$ in \eqref{scatdatafull},  taking $\Upsilon(i\kappa)=0, \, \kappa>0$, and with a little bit of algebra, one could separate those terms involving an exponential from those with polynomials,  obtaining the result given on page 7 (eqs.  \eqref{spec1} and \eqref{spectralfull}).

The Green's function of the associated QFT can be expressed in a compact way as \begin{equation}
G_{k}(z_1, z_2)=G_{k}^{PT}(z_1, z_2)+ \Delta G_{k}(z_1, z_2),
\end{equation} 
with $\Delta G_{k}(z_1, z_2)$  given by \eqref{Green1}. 
\small
\begin{eqnarray}\label{Green1}
\left\{\begin{array}{lll}
 \frac{r_L}{W} f_{k}(z_1)f_{k}(z_2), & \qquad \qquad \qquad & \textrm{if}\, \, z_1, z_2>b,\\[3ex]
 \frac{r_R}{W} f_{-k}(z_1)f_{-k}(z_2), & & \textrm{if}\, \, z_1, z_2<a,\\[3ex]
 \frac{B_R B_L}{t\, W} f_{k}(z_1)f_{k}(z_2) + \frac{C_R C_L}{t\, W} f_{-k}(z_1)f_{-k}(z_2) & & \textrm{if}\, \, a<z_1<b\,\, \textrm{and}\,  \,  a<z_2<b,\\[1ex]
+ \frac{C_R B_L}{t\, W} (f_{-k}(z_>)f_{k}(z_<)+f_{k}(z_>)f_{-k}(z_<)), & &\\[3ex]
(t-1) \, G^{PT}_k (z_1, z_2),  & & \textrm{if}\, \, z_2>b\,\, \textrm{and}\,  \, z_1 <a \, \textrm{(or $z_1\leftrightarrow z_2$)},\\[3ex]
(C_L-1) \, G^{PT}_k (z_1, z_2) + \frac{B_L}{W} f_{k}(z_1)f_{k}(z_2), & & \textrm{if}\, \, z_2 >b\,\, \textrm{and}\,  \,  a<z_1 <b\, \textrm{(or $z_1 \leftrightarrow z_2$)},\\[3ex]
(B_R-1) \, G^{PT}_k (z_1, z_2) + \frac{C_R}{W} f_{-k}(z_1)f_{-k}(z_2), & & \textrm{if}\, \, z_2 <a\,\, \textrm{and}\,  \,  a<z_1 <b \, \textrm{(or $z_1\leftrightarrow z_2$).}
\end{array}\right. 
\end{eqnarray}
\normalsize
The scattering data involved are collected in \eqref{scatdatafull}. The points $\{a,b\}$ at which the plates are located are completely general and can be replaced by any other pair of points, independently of their position with respect to the origin, around which the PT kink is centered.

\section{ Proof of relation \eqref{TGTGAPROX}}
\label{AppendixB}
In Appendices B and C of \cite{KK2008} the authors prove that if the Green's function $G(x,y)$ is smooth for $x\neq y$, then for any two disjoint objects $A,B$ separated by a finite distance, $G^{AB}$ is a trace-class operator. Moreover, $T^A$ and $T^B$ are bounded and $T^A G^{AB}T^BG^{BA}$ is a trace-class operator. They also prove that the modulus of the eigenvalues of $T^A G^{AB}T^BG^{BA}$ is less than one. The aim of this appendix is to demonstrate that
\begin{eqnarray}\label{APP1}
\tr \log(1-TGTG) &=&\log \, \det(1-TGTG)\approx  \log(1-\tr\, TGTG)
\end{eqnarray}
is satisfied whenever there are two separate objects in a weak curved background for which the aforementioned hypothesis for the $TGTG$ operator are satisfied, and $\textrm{rank}\, (TGTG)\neq 1$ is fulfilled. 

The first equality in \eqref{APP1} is general, and it can be proven by taking into account that any Hermitian matrix $P$ representing an Hermitian operator can be transformed into a diagonal matrix $P_D$, so that $P_D=Q P Q^{-1}$. In this way:
\begin{eqnarray}
&&e^{\textstyle \tr \log\, P}  = e^{ \textstyle\tr \log\, (Q^{-1} P_D Q)} = e^{ \textstyle \tr\, [Q^{-1} (\log P_D) Q]} = e^{\textstyle  \tr \log \, P_D} = e^{ \textstyle \sum_i \log \lambda_i} \nonumber\\
&&=\prod_i \lambda_i=\textrm{det}\,  P_D= \textrm{det}\,  (Q P Q^{-1})=  \textrm{det}\, P.
\end{eqnarray}
The cyclic property of the trace has been used above. The eigenvalues of the diagonal matrix $P_D$ have been denoted by $\lambda_i$. On the other hand,  if $\alpha_i$ are the eigenvalues of $M=TGTG$, it is easy to prove the second claim in \eqref{APP1},  because
\begin{eqnarray}
&&\log \textrm{det}\, (1-M) =  \log \textrm{det}\,[Q_1^{-1} (1-M_D) Q_1]   = \log \prod_i (1-\alpha_i) = \log [1- \sum_i \alpha_i+o(\alpha_i \alpha_j)]\nonumber\\
&&\approx   \log [1-\tr\, M_D] = \log [1-\tr ( Q_1 M Q_1^{-1})] =\log [1-\tr \, M ].
\end{eqnarray}
It constitutes a good approximation up to first order since the norm of $M$ is less than one.

In some references, another approach to understand the approximate computation can be found. If we consider dielectrics instead of the vacuum, if the dielectric function can be written as $1+x $ being $x$ small (i.e., the dilute limit), $||TGTG||<1$,  and one could make the formal expansion of $\log(1-x)$ to finally write  \begin{equation}
\log \det (1-TGTG) =-\sum_j \frac{1}{j} \textrm{Tr} \, (TGTG)^j
\end{equation} 
(this is done in Sec. V of the seminal paper of Kenneth and Klich \cite{KK2008} or in SEC. 4.1 of \cite{Bordag2012}). In this last article, figure 4 shows the numerical difference between the contribution of the integral \eqref{KK2008for} for $j=1$ and $j=2$.  As can be seen in the left plot of this figure, considering  $\textrm{Tr} \, (TGTG)^2$ gives a result numerically much smaller than the one involving $\textrm{Tr} \, (TGTG)$,  so we can neglect it and still obtain a good approximation for the quantum vacuum correction to the Casimir energy up to first order in $\hbar$). These type of approximations are widely used in the literature, for instance in \cite{Guilarte2013, Guilarte2015}.

There is another important detail to be considered. Notice that \begin{equation}
\textrm{tr} \log (1 -TGTG) = \log \det (1-TGTG) = c_{TGTG}(1)
\end{equation} is always fulfilled.  $c_{TGTG}(1)$ is the characteristic polynomial of the matrix representing the operator $TGTG$. This polynomial can be expressed in terms of the trace of the operator itself and the trace of matrices given by products of a finite number of times the operator as:
\footnotesize
\begin{equation}\label{cdef}
c_{TGTG}(1)= 1+ \sum_{k=1}^n \frac{(-1)^k}{k!} \left| \begin{array}{cccccc}
\tr TGTG & 1 & 0 &0 & \hdots & 0\\
\tr (TGTG)^2 & \tr (TGTG) & 2 &0 & \hdots & 0\\
\vdots &&&&& \vdots\\
\tr (TGTG)^{k-1} & \tr (TGTG)^{k-2} & \tr (TGTG)^{k-3} & \tr (TGTG)^{k-4} & \hdots & k-1 \\
\tr (TGTG)^{k} & \tr (TGTG)^{k-1} & \tr (TGTG)^{k-2} & \tr (TGTG)^{k-3} & \hdots & \tr TGTG 
\end{array}\right|.
\end{equation}
\normalsize
 So the correct way to compute the Casimir energy between objects and obtain the exact result when $\textrm{rank} \, TGTG \neq 1$ is to evaluate numerically the following whole integral: 
\begin{eqnarray}
&&\int_m^\infty \xi \sqrt{\xi^2-m^2} \log \det (1-TGTG) 
= \int_m^\infty \xi \sqrt{\xi^2-m^2} \, \, c_{TGTG}(1).
\end{eqnarray}

\end{document}